# The effects of cavity on the etendue of a light source

Ui-Hyung Lee[1], Young-Gu Ju[2*]

[1] Department of Physics, Kyungpook National University, 1370 Sankyuk-dong, Buk-gu, Daegu, 702-701, Korea

[2] Department of Physics Education, Kyungpook National University, 1370 Sankyuk-dong, Buk-gu, Daegu, 702-701, Korea

[*] ygju@knu.ac.kr

## Abstract

The simulation proposes that the cylindrical cavity around a circular light source can decreases the divergence angle without changing the emission energy and source size. This result can provide a way to reduce the etendue of the light source by scrambling the rays inside the cavity with a lossless scattering surface. The experiment demonstrates that the metallic cavity around the surface source reduces the divergence angle. However, the metallic surface also absorbs quite a large portion of the light energy. The Lambertian nature of the sidewall surface changes the directions of the rays into horizontal directions. It increases the number of reflection inside the cavity and amplifies the small amount of absorption at a single reflection. The effect of the cavity on the etendue of the light source can contributes to providing more design flexibility in various lighting applications.

# 1. Introduction

One of the most frequently encountered physical theorems in lighting design is the etendue theorem. This theorem is well explained in most textbooks on illumination, usually at the beginning[1-3]. For instance, suppose that a designer is given a problem to maximize the illuminance of a headlight using a lens. The situation is illustrated in Fig. 1. If the source has a linear dimension of s and a divergence angle of $\omega$, then the product $s\omega$ never decreases in the image plane, no matter what kind of optics are used. From the image's viewpoint, the divergence angle is $\Omega$ and the image size is S. Since $\Omega$ is (a/b) multiplied by $\omega$ and S is (b/a) multiplied by s, the product $S\Omega$ remains same as $s\omega$. If the size s is replaced by the source area and $\omega$ by the projected solid angle, the product still remains unchanged. This product can decrease when the angle subtended by the edge of the lens is smaller than the divergence of the source. However, in this case, the lens fails to collect the light energy which passes outside the edge.

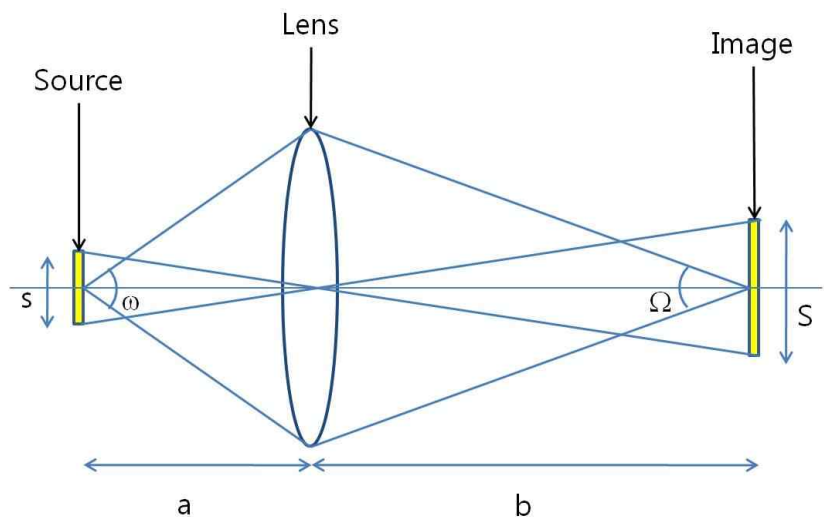

Fig. 1 Schematic diagram for illustrating etendue conservation in lighting applications

In this manner, the etendue theorem sets the lower limit of the image size since the angle $\Omega$ is usually fixed in a system such as a headlight. The distance between the lens and the screen is nominally fixed and the diameter of the lens has an upper limit in the system. The constraints on these two parameters give a certain number to $\Omega$. The consequence of the etendue theorem implies that the etendue of a light source already puts a lower bound on the size of the image and an upper bound on the maximum illuminance of the spot, what kind of optics a designer chooses. With a smaller etendue, a smaller spot and higher illuminance can be achieved. The situation doesn't change whether a designer adopts a single lens or double lens system. The etendue theorem claims that the external optics never decrease the etendue of the light source.

In this paper, we present a special circumstance where the etendue of a light source can decrease using an ideal cavity structure. The simulation will demonstrate that the divergence angle of the surface light source inside a box can decrease without the loss of source size and energy, resulting in reduced etendue. The experimental results also present that the light source within a cavity can have a decreased divergence angle with some loss of energy, which can be compensated by the gain of the reduced etendue and higher illuminance of the spot.

## 2. Simulation

The simulation is carried out in order to see if the divergence angle of the light source can be reduced without energy loss. LightTools 7.1 is used to compute the three different situations shown in Fig. 2. The three different models have the same light source, but have the cylindrical boxes with different heights. The first model consists of a simple circular light source with a diameter of 6 mm and a thickness of 0.01 mm. The circular light source generates a Lambertian intensity distribution. The second model combines the first circular light source and a cylindrical box with a height of 5 mm. The third model increases the height of the cylindrical box up to 10 mm to observe the effects of the height on the divergence angle. The inner diameter of the cylindrical box is equal to the diameter of the circular light source. The inner surface of the box is programmed to reflect and scatter the light with a Lambertian distribution without any energy loss. Although a perfect reflection with complete energy conservation is difficult to implement with a metal surface due to its absorption, this simulation helps us to understand the effects of cavity on the etendue in an ideal situation. Since the bottom of the cylindrical box also reflect the light perfectly, the light emitted from the source bounces back and forth inside the cavity and is only able to escape through the opening of the cylindrical box.

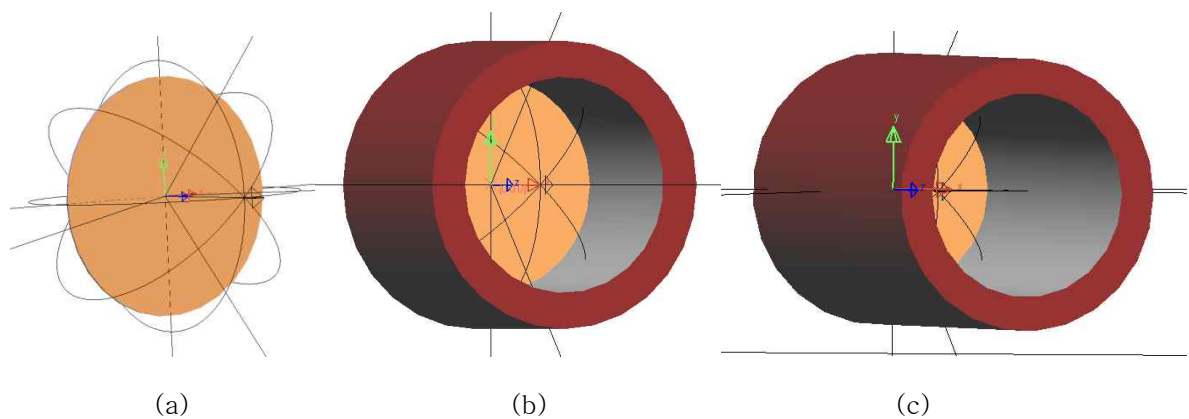

(a)　　　　　　　　　　(b)　　　　　　　　　　(c)

Fig. 2. (a) a circular light source   (b) a source and a cylindrical box with the height of 5 mm
(c) a source and a cylindrical box with the height of 10 mm.

The measurement of the divergence angle was done by placing a detector 300 mm from the source. In the calculation, the output power of the light source is set to 0.1 W. The final output power of the source is also obtained by integrating the light output detected along all directions in order to confirm the energy conservation. As expected, since the reflectance of the inner surface is set to 100 %, the total integrated power turns out to be 0.1 W.

The simulation shows that the divergence angle apparently decreases depending on the height of the cylindrical box as seen in Fig. 3. As the height of the wall increases, the divergence angle decreases. The divergence angle can be defined as the angle at which the intensity is half of the maximum intensity. The divergence angle of the Lambertian source is about 60 degree as shown in Fig. 3-(a). These angles become 45 degrees and 30 degrees for heights of 5 mm and 10 mm, respectively. This angle corresponds to the inverse tangent of the ratio of the diameter to the height of the cylindrical cavity. This is simply a three dimensional geometrical effect only if the circular source is replaced by the sum of the circular source and the side walls of the cavity. As the viewing angle deviates from the surface normal direction, the emittance of the source, including the sidewall, is blocked by the structure of the cavity.

Evidently, the size of the final source is the same as that of the cavity opening, which is the same as that of the first light source. As a result, the divergence angle decreases while the source size remains constant. This implies that the etendue of the source can decrease due to the cavity around the light source. In fact, it is controversial whether the cavity structure belongs to the external optics stated in etendue theorem. Optics such as lens and mirror usually deform the wavefront. However, the cavity in the above circumstances scatters the light and regenerates the light source over the side walls with an almost random sequence of rays. Anyway, this result is of practical use when someone wants to reduce the divergence angle of a surface source like an LED without increasing the source size. Relating to Fig. 1, the small divergence of the source enables a designer to select a lens with a longer focal length for a given diameter, which leads to a small spot size at the image plane with higher illuminance.

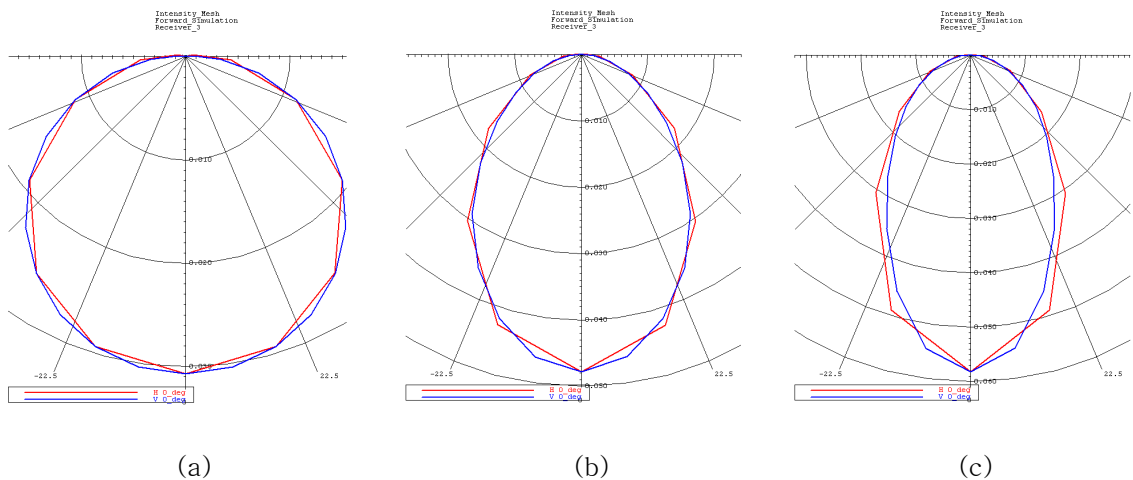

(a) (b) (c)

Fig. 3. Angular distribution of luminous intensity (a) a circular light source (b) a source and a cylindrical wall with the height of 5 mm (c) a source and a cylindrical wall with the height of 10 mm.

According to the additional simulations with less reflectivity of the inner surface, the total energy coming out of the cavity drops significantly. However, the angular intensity distribution remains unchanged since it is purely a geometrical effect. A discussion of the imperfect reflection of the cavity surface and the experimental results will be addressed in in the next section.

## 3. Experiments

The experimental setup is prepared in order to test if the divergence angle is reduced by the cavity structure and how much energy is conserved in the real structure. The light source consists of an acrylic cylinder with a diameter of 6 mm and the reflecting sheet on top of it. A 3M CM592 is used as a reflecting sheet and transmits the blue excitation light and reflects the red light. The reflecting sheet is cut in the shape of a circle matching the size of the acrylic post. The reflecting sheet is glued to the acrylic post using UV resin. The red phosphor is sprayed uniformly over the reflecting sheet. The cylindrical wall is realized by drilling a 6 mm hole into an aluminum block. The LED emitting at 460 nm illuminates the phosphor through the acrylic post and the reflecting sheet. In this manner, the fluorescent light from the red phosphor is reflected from the bottom reflecting sheet and the sidewalls of the aluminum block. The detector is placed 300 mm apart from the source and is equipped with the three red filters placed in front of the detector window. A red filter blocks the blue light and transmits most of red light with an extinction ratio greater than 1000. The detector is fastened to the rotating arm, which takes a measurement every 10 degrees from the normal direction.

The angular intensity distribution is plotted in Fig. 4. In the case of (a), the intensity at the surface normal direction is 176.4 nW and the divergence half angle is 60°, which agrees with the nominal Lambertian source. The cavity structure reduces the divergence angles as shown in Fig. 4-(b) and (c) depending on the heights of the cylindrical walls. In the case of (b), the intensity at surface normal direction is 200.8 nW and the divergence angle is 50°. The total integrated energy is 70% of that of the case (a). The total integrated energy is semi-empirically obtained by integrating the product of the angular distribution and a factor of $\sin\theta$ over the range of 90 degrees. Since the energy conservation ratio is important, the integrated power is normalized to 1 in the case of (a). The integrated powers of the other cases are described only as a ratio relative to the case of (a). When the height becomes 10 mm, the normal intensity reaches 187.1 nW and the divergence angle is 30°. In this structure, 42% of the energy is conserved with respect to the case (a).

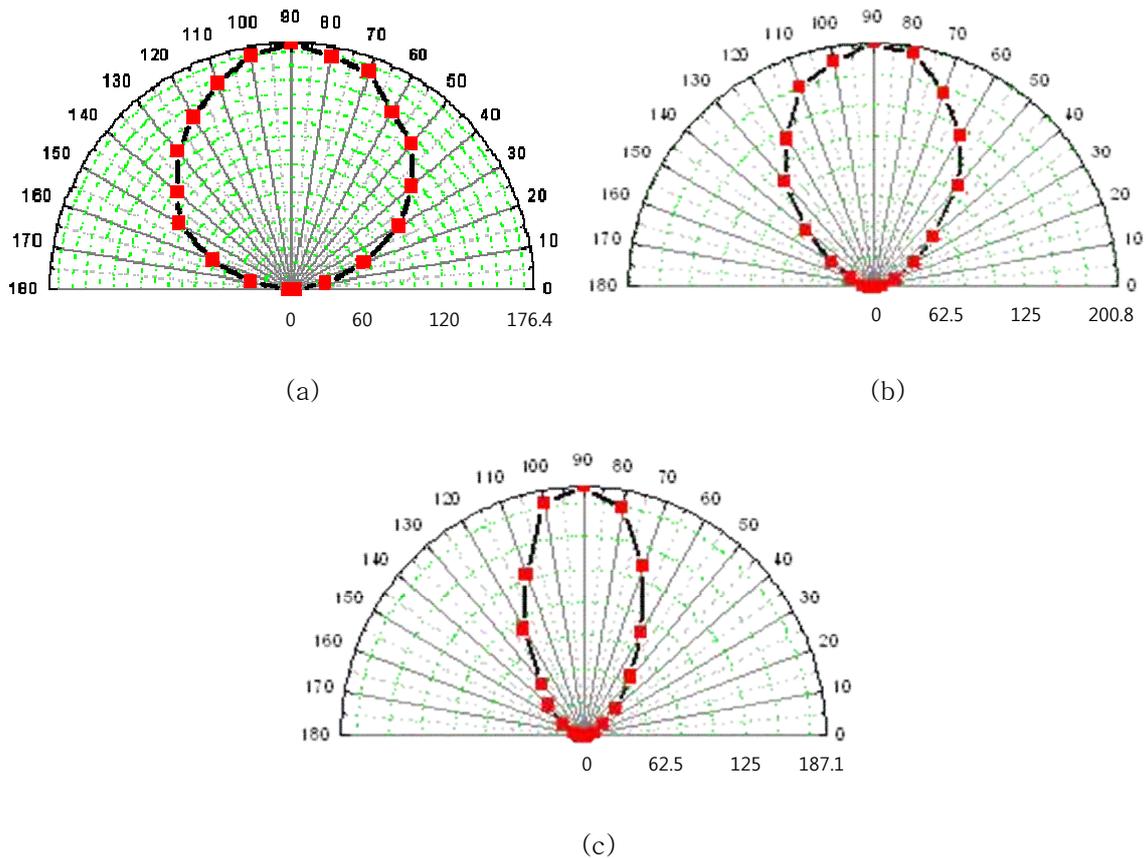

Fig. 4. Angular distribution of the luminous intensity is measured when phosphor is coated on a CM592 reflecting sheet. (a) a circular light source  (b) a source and a cylindrical wall with the height of 5 mm (c) a source and a cylindrical wall with the height of 10 mm.

The experiments demonstrate that the divergence angle decreases with the increase of the height of side walls. However, the total energy coming out of the cavity also diminishes significantly with increases in the height of the structure. This phenomenon can be explained by looking into the ray patterns inside the cavity. The traces of 50 rays inside the cavity are displayed in Fig. 5. The ray patterns show that more rays bounces back and forth horizontally rather than vertically. This trend becomes more serious when the aspect ratio of the height to the diameter increases. It can be ascribed to the scattering property of the surface. The Lambertian surface reflects the incident light into the direction normal to the surface instead of the specular angle. Therefore, the ray incident on the sidewall at finite angle, changes into the horizontal ones, which makes it more difficult to escape the cavity. This trapping phenomenon becomes serious in high sidewalls and results in a large energy loss if the surface at the wall absorbs the light energy during every reflection. Since the number of reflection in horizontal rays increases rapidly with the height of the structure, the small amount of energy loss at a single reflection becomes a problem.

The effects of the finite reflectivity on the energy efficiency are also investigated using simulation. In the model using the cylindrical box with the height of 10 mm, the reflectivity of the surface is assumed to be 90%, which is close to that of aluminum surface. The calculation shows that the energy coming out of the opening decreases to 48 % of a perfect reflection. It is in good agreement with the experimental data.

Although only 40% of the initial energy comes out of the cavity, it decreases the divergence angle by a half. It means that the designer chooses the lens with a focal length that is as twice as long. This will shrink the size of the image by half and the area by one fourth. In this case, the power per unit area or the illuminance doubles. Further research is required to increase the reflectivity of the cavity surface by choosing the appropriate material and, finally, to obtain the reduction of the etendue of the source without the sacrifice of energy.

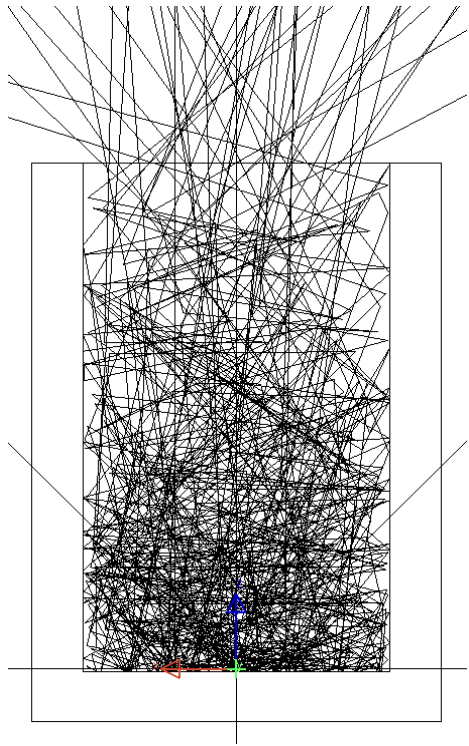

Fig. 5. Traces of 50 rays inside the 10 mm cylinder with a light source at the bottom.

## 4. Conclusion

The simulation proposes that the cylindrical cavity around a circular light source can decreases the divergence angle without changing the emission energy and source size. This result can provide a way to reduce the etendue of the light source by scrambling the rays inside the cavity with a lossless scattering surface. The experiment demonstrates that the metallic cavity around the surface source reduces the divergence angle, which is expected from the simulation. However, the metallic surface also absorbs quite a large portion of the light energy. The Lambertian nature of the sidewall surface changes the directions of the rays into horizontal directions. It increases the number of reflection inside the cavity and amplifies the small amount of absorption at a single reflection. With an appropriate choice of material for the side wall surface, a large portion of energy can be conserved with a reduced etendue of the light source. The effect of the cavity on the etendue of the light source can contributes to more design flexibility in various lighting applications.